Quark Mass Ratios in ChPT with the Difference of Quark Condensates Considered [*]

Xiaoya Li,[1]    Xiaofu Lü[1,2,3]

[1] Department of Physics, Sichuan University, Chengdu 610064, China

[2] Institute of Theoretical Physics, Chinese Academy of Science, Beijing 100080, China

[3] China Center of Advanced Science and Technology (World Laboratory), Beijing 100080, China

**Abstract**: The chiral effective Lagrangian for pseudoscalar nonet is constructed in consideration of isospin breaking. And the difference of quark condensates is taken accounted. The SU(3) singlet $\eta_0$ is not taken as Goldstone-boson. The $\eta$ mixing with and without isospin symmetry is considered. The quark mass ratios are obtained through solving the mass equations of mesons. We estimate the change of quark mass ratios according to the change of $M_{\pi^\pm}$ and $M_{K^\pm}$ to see how the electromagnetic corrections affect our results. It turn out that massless up quark is possible. The upper limit for $m_u/m_d$ is around 0.39. $2m_s/(m_u+m_d)$ =24.23~25.12. The values for $M_{\eta 0}$, quark condensate ratios and other constants are limited in narrow ranges.

**PACS numbers**: 12.39.Fe, 14.65.Bt, 14.40.Aq

**Keywords**: quark mass ratios, quark condensate, $\eta$ mixing, chiral perturbation theory

# 1    Introduction

Quark mass ratios are fundamental parameters in QCD, and need to be accurately determined. CP violation suggests that up quark might be massless [1] while previous work in chiral perturbation theory showed that $m_u/m_d \approx 0.553$ [2]. The data 0.3~0.7 of this ratio at the PDG is a summary of the extractions derived from chiral symmetry [3]. However, the ChPT estimation does not involve the difference among quark condensates of different flavors at leading order. In this paper, this difference is considered and the quark mass ratios are estimated in ChPT.

In chiral perturbation theory, the quark mass ratio can be derived through meson mass dependence of quark masses[2]. In order to reveal the u d quark difference, isospin breaking, i.e. $m_u \neq m_d$, must be considered. But isospin breaking leads not only to different u, d masses, but also to different u d quark condensates ($<\bar{u}u>, <\bar{d}d>$). Previous work in this field ignored the latter and the mass expressions of $\pi^0$ and $\pi^\pm$ turned out to be the same[2]. It was explained that the difference between the pionic masses was exclusively generated by electromagnetic interaction, which means that the isospin breaking has no effect on mesonic mass in QCD sector. It is unreasonable. The electromagnetic (EM) corrections for pions and kaons were discussed in Ref. [4], where the correction for pions was comparable to the experimental mass difference. But the result has an uncertainty of 30%.

---

[*] The project is supported in part by Key Research Plan of theoretical Physics and Cross Science of China under Grant No.90503011

We extend the method used in Ref. [5], where we have constructed the chiral effective Lagrangian for Goldstone bosons and $\eta_0$ in consideration of the difference between strange and non-strange quark condensate. Through solving the mass equations of the nonet pseudoscalar meson (the $\pi^0 - \eta_8 - \eta_0$ mixing was considered first), quark mass ratios and quark condensate ratioss can be determined. However, as the EM correction has not been determined yet, we can not directly use the physical value of meson masses to derive the result. The Dashen theorem indicates that [6], under chiral limit, the EM corrections for the square mass of charged meson $\pi^\pm$ and $K^\pm$ are the same while for neutral meson $\pi^0$ and $K^0$ vanish. So, we set an equal EM correction for $\pi^\pm$, $K^\pm$ to see how the quark mass ratios and quark condensate ratios response to the correction.

For $\eta$ mixing, we consider first the $\eta_8 - \eta_0$ mixing under isospin symmetry, and the mixing angle turned out to be $11.9^o$. When isospin symmetry is broken, $\pi^0$ is involved too. And the physical masses are the eingin values of the mixing mass matrix.

As for $\eta_0$, because of U(1)$_A$ breaking [7,8], it is not a Goldstone boson and should be processed solely instead of being processed together with the other eight members of Pseudoscalar meson neont.

This paper is organized as follows. In section 2, the effective lagrangian for Goldstone bosons and the SU(3) singlet $\eta_0$ is constructed in ChPT with explicit isospin breaking. In section 3, the mass expressions for pseudoscalar mesons are obtained. The $\eta$ mixing with and without isospin symmetry is considered. Setting equal EM correction, the quark mass ratios and quark condensate ratios are estimated through solving the mass equations of mesons. Section 4 gives the conclusion.

## 2  Effective Lagrangian in ChPT

### 2.1  Introduction of Quark External Fields

In order to reveal the quark information, our effective Lagrangian involves the external fields connecting with quarks in the following way. The external fields $v_\mu, a_\mu, s, p$ are defined as [9]

$$\ell_{ext}^{QCD} = \ell_0^{QCD} + \bar{q}\gamma^\mu\{v_\mu + \gamma_5 a_\mu\}q - \bar{q}\{s - i\gamma_5 p\}q, \qquad (1)$$

Where $\ell_0^{QCD}$ denotes the QCD Lagrangian under chiral limit (m$_u$=m$_d$=m$_s$=0). $v_\mu, a_\mu, p, s$ are hermitian matrices and $tra_\mu = trv_\mu = trp = 0$. Under $SU(3)_L \times SU(3)_R$ they transform as

$$v_\mu + a_\mu \to V_R(v_\mu + a_\mu)V_R^+, \qquad (2)$$

$$v_\mu - a_\mu \to V_L(v_\mu - a_\mu)V_L^+, \qquad (3)$$

$$s + ip \to V_R(s + ip)V_L^+, \qquad (4)$$

$$s - ip \to V_L(s - ip)V_R^+. \qquad (5)$$

Here we consider global chiral transformation other than local as was done in Ref. [9]. In real world, generating functional $\exp(iZ(v_\mu,a_\mu,s,p)) = \langle 0_{out}|0_{in}\rangle_{v,a,s,p}$ is expanded around $v_\mu = a_\mu = p = 0$, $s = diag(m_u, m_d, m_s)$. Generating functional are calculated effectively in a chiral perturbation way [9] as $\exp(iZ(v_\mu,a_\mu,s,p)) = \int [du][d\eta]...\exp(i\int d^4x \ell_{eff})$, where ... denotes all kinds of hadrons. When hadrons fields are integrated out in the path integral, only external fields remain. The physical vacuum expectation value of time-ordered operators $\bar{q}\gamma_\mu\lambda^a q$, $\bar{q}\gamma_\mu\gamma_5\lambda^a q$, $\bar{q}q$, $\bar{q}\gamma_5 q$, etc. are obtained by calculating the functional differentiation in generating to the corresponding external fields. At leading order, the generating functional is calculated at the classical solution of Lagrangian equation. i.e. U(x) is taken as the solution $U_0$ to the equations of motion [9].

## 2.2 Symmetry Transformation Properties of Pseudoscalar Mesons

The effective Lagrangian should contain all the possible terms consistent with the symmetry principles according to Weinberg [10]. At low energy regime, the calculation is done in a perturbation way [9-11]. So, calculating to a certain order, we need to consider only those Lagrangian terms having contribution to our calculation at this given order. To count the power, power counting was introduced by Weinberg [10] and had been used by Gasser and Leutwyler in their work at ChPT [9,11]. In this paper, we concentrate ourselves on $O(p^2)$ and consider all the $O(p^2)$ Lagrangian terms having contribution to the calculation of meson mass. In pseudoscalar meson system, the effective Lagrangian consists of Goldstone bosons and $\eta_0$.

Eight Goldstone bosons $\phi$ are contained in a unitary matrix $U(x) = \exp(i\phi/F)$. Under nonlinear realization of $SU(3)_L \times SU(3)_R$, U(x) transforms as [9]

$$U(x) \to V_R U(x) V_L^+. \tag{6}$$

Let $U(x) = u(x)^2$, then u(x) transforms like [9]

$$u(x)' = V_R u(x) R^{-1} = R u(x) V_L^{-1}, \tag{7}$$

Where

$$R = u'^{-1} V_R u = u' V_L u^{-1}. \tag{8}$$

For later convenience, we define two axial vector fields

$$u_\mu = -\frac{1}{2}(u^+ \partial_\mu u - u \partial_\mu u^+), \quad u_\mu \to R u_\mu R^+, \tag{9a}$$

$$A_\mu = \frac{i}{2}u^+(v_\mu + a_\mu)u - \frac{i}{2}u(v_\mu - a_\mu)u^+, \quad A_\mu \to R A_\mu R^+. \tag{9b}$$

Let $\chi = 2B_0(s+ip)$, where $B_0$ is a real constant with a dimension of mass, we define a scalar and a pseudoscalar matrix

$$\chi_+ = u\chi^+ u + u^+ \chi u^+, \tag{10a}$$

$$\chi_- = u\chi^+ u - u^+ \chi u^+, \tag{10b}$$

which transform as

$$\chi_{\pm} \to R\chi_{\pm}R^+ \tag{11}$$

under nonlinear realization of $SU(3)_L \times SU(3)_R$.

Because U(1)$_A$ breaking happens before spontaneous breaking of chiral symmetry [7], there are only eight Goldstone bosons and the SU(3) singlet $\eta_0$ is not a Goldstone boson. The singlet is invariant under nonlinear realization of $SU(3)_L \times SU(3)_R$

$$\eta_0' = \eta_0. \tag{12}$$

The quark condensates are introduced through the VEV of scalar meson $S = \bar{q}\lambda_a q \lambda_a$ [10], where $a = 0,1,2,\ldots 8$. Under nonlinear realization of $SU(3)_L \times SU(3)_R$, S transforms as

$$S \to RSR^+. \tag{13}$$

The only nonvanishing components of $<S>$ are those with a=0, 3, 8

$$<S> = <\bar{q}\lambda_0 q \lambda_0> + <\bar{q}\lambda_3 q \lambda_3> + <\bar{q}\lambda_8 q \lambda_8> = diag(<\bar{u}u>,<\bar{d}d>,<\bar{s}s>). \tag{14}$$

**2.3 Goldstone Bosons**

We define a matrix $X = diag(\sigma_1, \sigma_2, \sigma_3)$, where $\sigma_1 \sim <\bar{u}u>$, $\sigma_2 \sim <\bar{d}d>$, $\sigma_3 \sim <\bar{s}s>$. Similar to the isospin symmetric form introduced in Ref. [5,12], the kinetic energy term of Goldstone bosons is

$$Tr[Xu_\mu X u^\mu + X u_\mu u^\mu X]. \tag{15}$$

To obtain the explicit kinetic energy terms, $\phi/F$ must be defined as

$$\phi/F = \begin{pmatrix} \dfrac{\pi^0}{\sqrt{2}\sigma_1} + \dfrac{\eta_8}{\sqrt{6}\sigma_1} & \dfrac{2\pi^+}{\sigma_1+\sigma_2} & \dfrac{2K^+}{\sigma_1+\sigma_3} \\ \dfrac{2\pi^-}{\sigma_1+\sigma_2} & -\dfrac{\pi^0}{\sqrt{2}\sigma_2} + \dfrac{\eta_8}{\sqrt{6}\sigma_2} & \dfrac{2K^0}{\sigma_2+\sigma_3} \\ \dfrac{2K^-}{\sigma_1+\sigma_3} & \dfrac{2\bar{K}^0}{\sigma_2+\sigma_3} & -\sqrt{\dfrac{2}{3}}\dfrac{\eta_8}{\sigma_3} \end{pmatrix}. \tag{16}$$

Since $A_\mu$ and $u_\mu$ have identical symmetry transformation properties, the following terms are allowed

$$\frac{1}{2}Tr[Xu_\mu X A^\mu + X A_\mu X u^\mu + X u_\mu A^\mu X + X A_\mu u^\mu X], \tag{17}$$

which contains the coupling of Goldstone bosons to the quark external field $a_\mu$. Strictly speaking, the Lagrangian term in Eq. (17) should be multiplied by a dimensionless constant. But it is easy to see that such a multiplier can be counteracted by $\sigma_i$'s. So, we can just define the multiplier as 1.

In real world, $s = diag(m_u, m_d, m_s)$ explicitly breaks chiral symmetry and makes the Goldstone bosons massive. At O(p$^2$), the explicit breaking term of the effective Lagrangian containing only Goldstone bosons is

$$\frac{1}{2}Tr[X\chi_+ X]. \tag{18}$$

Compared with the kinetic term $F_0^2 Tr[\partial_\mu U^+ \partial_\mu U]/4 = Tr[u_\mu u^\mu]$ and the explicit breaking term $F_0^2 Tr[\chi U^+ + U\chi^+]/4 = F_0^2 Tr[\chi_+]/4$ in Ref. [2], it can be seen that in our theory all the $F_0$ is replaced by a matrix $\sqrt{2}X$. Solving the motion equations of U(x), we find that our solution is $U_0 = F_0/X$ rather than $U_0 = 1$ as was given in Ref. [2]. With the solution obtained, vacuum quark condensate could be calculated. $\langle 0|\bar{q}q|0\rangle$ is calculated through deriving the generating functional with the external field s

$$\langle \bar{q}q\rangle = \frac{\delta}{\delta s} Z|_{v_\mu = a_\mu = p, s = diag(m_u, m_d, m_s)}. \tag{19}$$

At leading order, our theory shows that

$$\langle \bar{q}\lambda^a q\rangle = B_0 F_0 Tr[X\lambda^a], \tag{20}$$

$$\langle \bar{u}u\rangle = B_0 F_0 \sigma_1, \quad \langle \bar{d}d\rangle = B_0 F_0 \sigma_2, \quad \langle \bar{s}s\rangle = B_0 F_0 \sigma_3, \tag{21}$$

which is consistent with our assumption at the beginning of this section. In Ref. [2], the leading order result is $\langle \bar{u}u\rangle = \langle \bar{d}d\rangle = \langle \bar{s}s\rangle = B_0 F_0^2$, which did not reveal the quark condensate difference.

Following the method used in Ref. [2], the decay constants of Goldstone bosons are obtained at leading order through

$$\langle 0|\bar{q}\gamma_\mu \gamma_5 \lambda^a q|b\rangle = ip_\mu F^{ab}, \tag{22}$$

Where $|b\rangle$ denotes the Goldstone boson states and a, b are the SU(3) generator indices. When isospin symmetry is maintained, the results turn out to be

$$F_\pi = -\frac{\sigma_1 + \sigma_2}{\sqrt{2}}, \tag{23a}$$

$$F_K = -\frac{\sigma_1 + \sigma_2 + 2\sigma_3}{2\sqrt{2}}. \tag{23b}$$

$$F_\eta = -\frac{\sigma_1 + \sigma_2 + 4\sigma_3}{3\sqrt{2}} \tag{23c}$$

Using the experimental value of $F_K/F_\pi = 1.22 \pm 0.02$ [2,3], we get $2\sigma_3/\sigma_1 + \sigma_2 = 1.44 \pm 0.04$. The one-loop calculation of $F_\eta/F_\pi$ is $1.3 \pm 0.05$ in Ref. [2], then $2\sigma_3/(\sigma_1 + \sigma_2) = 1.45 \pm 0.075$. It is glad to see that the two results derived from our equations are consistent. So, the leading order relationship between decay constant and quark condensate revealed in Eq. (23a-c) is suitable. In later calculation, we take $2\sigma_3/(\sigma_1 + \sigma_2)$, i.e. $2\langle\bar{s}s\rangle/(\langle\bar{u}u\rangle + \langle\bar{d}d\rangle)$, as 1.40~1.48. The condensate of strange quark is larger than that of up and down quarks in our result.

**2.3** $\eta_0$

As a singlet in SU(3)$_f$, $\eta_0$ is invariant under nonlinear realization of chiral symmetry. Its kinetic term is

$$-\frac{1}{2}(\partial_\mu \eta_0)^2. \tag{24}$$

Since $\eta_0$ is not a Goldstone boson, it has a zero order mass term, which denotes the $\eta_0$ mass under chiral limit

$$-\frac{1}{2}m_0^2 \eta_0^2. \tag{25}$$

This $m_0$ takes contribution from gluon condensate, which violates chiral symmetry and leads to a heavy $\eta_0$.

There are two explicit breaking terms

$$-\frac{1}{2}g_1 Tr[\chi_+]\eta_0^2, \tag{26a}$$

$$\frac{i}{2}g_2 Tr[X\chi_-]\eta_0. \tag{26b}$$

Eq. (26a) contains the O(p2) mass contribution from explicit breaking. Eq. (26b) contains the mixing between $\eta_0$, $\eta_8$ and $\pi_0$. Those terms in which $\eta_0$ has a power larger than 2 are allowed in principle but contribute only to multi-meson interactions. $\chi_+$ exists in those terms where even power of $\eta_0$ is involved while $\chi_-$ corresponds to odd power of $\eta_0$.

**3 Masses of Psudoscalar Mesons**

**3.1 Mass Expressions**

Expanding u(x) at $\phi$, we can get Goldstone boson masses from Eq. (18) and $\eta_0$ mass from Eq. (25) and Eq. (26a) at leading order.

$$M_{\pi\pm}^2 = \frac{2B_0(m_u+m_d)(\sigma_1^2+\sigma_2^2)}{(\sigma_1+\sigma_2)^2} \tag{27a}$$

$$M_{K\pm}^2 = \frac{2B_0(m_u+m_s)(\sigma_1^2+\sigma_3^2)}{(\sigma_1+\sigma_3)^2} \tag{27b}$$

$$M_{K0}^2 = \frac{2B_0(m_d+m_s)(\sigma_2^2+\sigma_3^2)}{(\sigma_2+\sigma_3)^2} \tag{27c}$$

$$M_{\pi 0}^2 = B_0(m_u+m_d) \tag{27d}$$

$$M_{\eta 8}^2 = \frac{1}{3}B_0(m_u+m_d+4m_s) \tag{27e}$$

$$M_{\eta 0}^2 = m_0^2 + 4B_0 g_1(m_u+m_d+m_s) \tag{27f}$$

In above equations, $M_{\pi 0}^2$ and $M_{\pi \pm}^2$ are distinguished. This is because the difference between $\sigma_1$ and $\sigma_2$ is considered, i.e. $<\bar{u}u> \neq <\bar{d}d>$. If the difference between quark condensates is ignored, our results coincide with expressions given in Ref. [2]. $M_{\eta 0}$ contains a zero order mass and a O(p$^2$) contribution.

Eq. (18) and Eq. (25b) contain the mixing terms $-A_{\pi 0\eta 8}\pi^0\eta^8$, $-A_{\pi 0\eta 0}\pi^0\eta^0$, $-A_{\eta 8\eta 0}\eta^8\eta^0$, where

$$A_{\pi 0\eta 8} = \frac{B0(-md+mu)}{\sqrt{3}}, \tag{28a}$$

$$A_{\pi 0\eta 0} = \sqrt{2}B_0 g_2 (m_u - m_d), \tag{28b}$$

$$A_{\eta 8\eta 0} = -\sqrt{\frac{2}{3}} B_0 g_2 (-2m_s + m_d + m_u). \tag{28c}$$

The $\pi^0 - \eta_8 - \eta_0$ mixing mass matrix is

$$Mx = \begin{pmatrix} M_{\pi 0}^2 & A_{\pi 0\eta 8} & A_{\pi 0\eta 0} \\ A_{\pi 0\eta 8} & M_{\eta 8}^2 & A_{\eta 8\eta 0} \\ A_{\pi 0\eta 0} & A_{\eta 8\eta 0} & M_{\eta 0}^2 \end{pmatrix}. \tag{29}$$

When isospin is maintained, the mixing mass matrix turns into a $2 \times 2$ matrix

$$Mxi = \begin{pmatrix} M_{\eta 8}^2 & A_{\eta 8\eta 0} \\ A_{\eta 8\eta 0} & M_{\eta 0}^2 \end{pmatrix}. \tag{30}$$

### 3.2 $\eta$ Mixing

The terms $-A_{\pi 0\eta 8}\pi^0\eta^8$, $-A_{\pi 0\eta 0}\pi^0\eta^0$ and $-A_{\eta 8\eta 0}\eta^8\eta^0$ in the Lagrangian shows that $\pi^0$, $\eta_8$ and $\eta_0$ are mixed. The physical particles observed in experiments are results of the mixing, which are denoted as $\tilde{\pi}^0$, $\eta$ and $\eta'$. At O(p2), the masses of the physical particles are obtained through diagonalizing the mixing mass matrix.

Under isospin symmetry ($m_u/m_d = 1$, $\sigma_1/\sigma_2 = 1$), $g_2$ and $M_{\eta 0}$ are obtained through solving the secular equation of the mixing matrix Mxi $g_2 = \pm 0.33, M_{\eta 0} = 944.3 MeV$.*

If $\eta^8$ and $\eta^0$ are mixed into $\eta$ and $\eta'$ in a pattern that

$$\eta = \eta_8 Cos\theta + \eta_0 Sin\theta, \tag{31a}$$

$$\eta' = -\eta_0 Cos\theta + \eta_8 Sin\theta, \tag{31b}$$

Then

---

* We have let $2\sigma_3/(\sigma_1+\sigma_2) = 1.44$, $2m_s/(m_u+m_d) = 25$, $M_{\pi 0} = 136 MeV$, $M_\eta = 548 MeV$, $M_{\eta'} = 958 MeV$.

$$M_{R\eta 8}{}^2 = M_\eta{}^2 Cos^2\theta + M_{\eta'}{}^2 Sin^2\theta, \tag{32a}$$

$$M_{R\eta 0}{}^2 = M_{\eta'}{}^2 Cos^2\theta + M_\eta{}^2 Sin^2\theta, \tag{32b}$$

$$A_{\eta 8\eta 0} = (M_\eta{}^2 - M_{\eta'}{}^2) Sin\theta Cos\theta. \tag{32c}$$

The mixing angle $\theta$ is $\pm 11.9^o$. Its sign is consistent with that of $g_2$.

The results under isospin symmetry are shown in Table 1.

Table 1 The values of mixing angle, quark mass ratios, quark condensate ratios, $g_2$, $g_1$, $M_{\eta 0}$ and $m_0$ under isospin symmetry.

| $m_u/m_d$ | $2m_s/(m_u+m_d)$ | $\langle \bar{u}u \rangle / \langle \bar{d}d \rangle$ | $2\langle \bar{s}s \rangle/(\langle \bar{u}u \rangle + \langle \bar{d}d \rangle)$ | $\theta$ | $g_2$ | $M_{\eta 0}$/MeV |
|---|---|---|---|---|---|---|
| 1 | 25 | 1 | 1.44 | $\pm 11.9^O$ | $\pm 0.33$ | 944.3 |

### 3.3 Equation Solutions and Results under Isospin Breaking

According to section 3.1, the values for quark mass ratios, quark condensate ratios, $g_2$ and $M_{\eta 0}$ can be obtained through combining Eq. (27a~c) and the secular equation of mixing matrix Mx. However, as the EM correction is not negligible, we could not directly input the physical masses. According to Dashen theorem [6], the EM correction for $M_{K\pm}{}^2$ and $M_{\pi\pm}{}^2$ are the same, while for $M_{\pi 0}{}^2$ and $M_{K0}{}^2$ vanishes. We subtract an equal value $\tau$ from the square masses of charged mesons,

$$M_{K\pm}{}' = \sqrt{M_{K\pm}{}^2 - \tau}, \tag{33}$$

$$M_{\pi\pm}{}' = \sqrt{M_{\pi\pm}{}^2 - \tau}. \tag{34}$$

So in our mass equations we should use $M_{K\pm}{}'$ and $M_{\pi\pm}{}'$ instead of $M_{K\pm}$ and $M_{\pi\pm}$. Table 2 shows the results of solving of Eq. (27a~c) and the secular equation of Mx. If we take the value of $2\langle \bar{s}s \rangle/(\langle \bar{u}u \rangle + \langle \bar{d}d \rangle)$ to be around 1.44, the values of the constants are in an increasing or decreasing row with the increase of $m_u/m_d$ from 0 to 0.378. When $m_u/m_d$ goes beyond 0.378, the equations have no solution at $2\langle \bar{s}s \rangle/(\langle \bar{u}u \rangle + \langle \bar{d}d \rangle) = 1.44$. To determine the range of the constants in Table 2, the error of $2\langle \bar{s}s \rangle/(\langle \bar{u}u \rangle + \langle \bar{d}d \rangle)$ should be taken accounted. As the values of those constants are increasing or decreasing regularly, we just need to apply the error at the lower and upper limit of $m_u/m_d$. The results are shown in Table 3. It can be seen that the upper limit for $m_u/m_d$ is around 0.39. The EM correction for charged mesons allowed in our theory is no more than 1198 MeV$^2$. The ranges for values of $2m_s/(m_u+m_d)$, $M_{\eta 0}$, $g_2$ and $\sigma_1/\sigma_2 = \langle \bar{u}u \rangle / \langle \bar{d}d \rangle$ are

$$2m_s/(m_u+m_d) = 24.23 \sim 25.12, \tag{35a}$$

$$M_{\eta 0} = 951.58 \sim 953.86 MeV, \tag{35b}$$

$$g_2 = \pm 0.19 \sim 0.24, \qquad (35c)$$

$$\langle \bar{u}u \rangle / \langle \bar{d}d \rangle = 0.681 \sim 0.923. \qquad (35d)$$

In our theory, massless up quark is possible. But in that case there is a large difference between up and down quark condensates.

Table 2 The values for quark mass ratios, quark condensate ratios, $g_2$ and $M_{\eta 0}$ under isospin breaking when $2\langle \bar{s}s \rangle / (\langle \bar{u}u \rangle + \langle \bar{d}d \rangle)$ is fixed around 1.44.

| $m_u/m_d$ | $\tau / MeV^2$ | $2\langle \bar{s}s \rangle / (\langle \bar{u}u \rangle + \langle \bar{d}d \rangle)$ | $\langle \bar{u}u \rangle / \langle \bar{d}d \rangle$ | $2m_s/(m_u + m_d)$ | $g_2$ | $M_{\eta 0}$/MeV |
|---|---|---|---|---|---|---|
| 0.00 | 255.0 | 1.441 | 0.698 | 24.31 | ±0.21 | 953.22 |
| 0.05 | 495.0 | 1.440 | 0.734 | 24.48 | ±0.21 | 952.97 |
| 0.10 | 685.0 | 1.441 | 0.768 | 24.60 | ±0.21 | 952.79 |
| 0.15 | 833.0 | 1.440 | 0.800 | 24.71 | ±0.22 | 952.64 |
| 0.20 | 950.0 | 1.440 | 0.830 | 24.80 | ±0.22 | 952.54 |
| 0.25 | 1041.6 | 1.440 | 0.857 | 24.87 | ±0.22 | 952.46 |
| 0.30 | 1113.0 | 1.441 | 0.884 | 24.93 | ±0.22 | 952.42 |
| 0.35 | 1168.0 | 1.441 | 0.908 | 24.97 | ±0.22 | 952.38 |
| 0.378 | 1192.8 | 1.441 | 0.922 | 24.99 | ±0.22 | 952.37 |

Table 3 The values for quark mass ratios, quark condensate ratios, $g_2$ and $M_{\eta 0}$ under isospin breaking when the error of $2\langle \bar{s}s \rangle / (\langle \bar{u}u \rangle + \langle \bar{d}d \rangle)$ is applied to the upper and lower limit of $m_u/m_d$.

| $m_u/m_d$ | $\tau / MeV^2$ | $2\langle \bar{s}s \rangle / (\langle \bar{u}u \rangle + \langle \bar{d}d \rangle)$ | $\langle \bar{u}u \rangle / \langle \bar{d}d \rangle$ | $2m_s/(m_u + m_d)$ | $g_2$ | $M_{\eta 0}$/MeV |
|---|---|---|---|---|---|---|
| 0.000 | 166 | 1.400 | 0.681 | 24.39 | ±0.22 | 952.63 |
| 0.000 | 325 | 1.480 | 0.712 | 24.23 | ±0.19 | 953.86 |
| 0.391 | 1198 | 1.400 | 0.923 | 25.12 | ±0.24 | 951.58 |
| 0.368 | 1189.7 | 1.479 | 0.919 | 24.87 | ±0.21 | 953.13 |

## 4  Conclusion

Quark mass ratios are essential parameters in QCD, but have not been well determined. Strong CP violation suggests a massless up quark [1] while previous ChPT estimation showed a large up quark mass [2]. The problem with the method used in Ref. [2] is that the difference of quark condensates is not considered at leading order. Taking account this difference, we constructed the chiral effective Lagrangian for pseudoscalar nonet. The relations between decay constants and quark condensates were obtained in Eq. (23a-c). Taking the experimental value and sum rule estimation for decay constants [9], $2\langle \bar{s}s \rangle / (\langle \bar{u}u \rangle + \langle \bar{d}d \rangle)$ turns out to be around 1.44. After treating the $\eta$ mixing, we solved the meson mass equations and estimated the quark mass ratios. Supposing that the EM correction for square masses of charged mesons are the same, we estimated the ranges for values of $2m_s/(m_u+m_d)$, $M_{\eta 0}$, $g_2$ and $\sigma_1/\sigma_2 = \langle \bar{u}u \rangle / \langle \bar{d}d \rangle$, which are shown in Eq. (35a-d). Our results indicate that massless up quark is possible and $m_u/m_d$ has an upper limit of 0.39. The EM correction for charged mesons allowed in our theory is no more than 1198 $\text{MeV}^2$.